# EU-28's progress towards the 2020 renewable energy share. A club convergence analysis

**María José Presno**[1]

**Department of Applied Economics**

**University of Oviedo**

**Avenida del Cristo, s/nº**

**33011 Oviedo**

**Asturias (Spain)**

**Phone: +34 985103758**

**Email: mpresno@uniovi.es**

**ORCID: 0000-0002-8691-4027**

**Manuel Landajo**

**Department of Applied Economics**

**University of Oviedo**

**Avenida del Cristo, s/nº**

**33011 Oviedo**

**Asturias (Spain)**

**Phone: +34 985105055**

**Email: landajo@uniovi.es**

**ORCID: 0000-0002-5357-955X**

---

[1] Corresponding author.

# EU-28's progress towards the 2020 renewable energy share. A club convergence analysis

**Abstract**

This paper assesses the convergence of the EU-28 countries towards their common goal of 20% in the renewable energy share indicator by year 2020. The potential presence of clubs of convergence towards different steady state equilibria is also analyzed from both the standpoints of *global* convergence to the 20% goal and *specific* convergence to the various targets assigned to Member States. Two clubs of convergence are detected in the former case, each corresponding to different RES targets. A probit model is also fitted with the aim of better understanding the determinants of club membership, that seemingly include real GDP per capita, expenditure on environmental protection, energy dependence, and nuclear capacity, with all of them having statistically significant effects. Finally, convergence is also analyzed separately for the transport, heating and cooling, and electricity sectors.

## 1. Introduction

The European Union (EU) has emphasized the crucial role of renewable energy sources (RES) in order to achieve the five dimensions of the Energy Union, namely security of energy supply (reducing dependence on fossil fuels and imported energy), market integration (by participating in markets on an equal footing with other energy sources), energy efficiency (reducing primary energy consumption and improving the energy performance of buildings), decarbonization (by reducing greenhouse emissions), and innovation (prioritizing research to further drive the energy transition). The progress in RES technologies also has a strong potential to boost Europe´s employment and economy, improve its air quality and industrial competitiveness (by reducing energy costs), and contribute to international development (by improving access of developing countries to affordable and clean energy sources).

The European Union has been one of the pioneers in the field of generation of renewable energy, with a 12% RES target by 2010 declared as early as in year 1997. Directive 2009/28/EC on the promotion of the use of energy from renewable sources (Renewable Energy Directive I, RED I) has been a central element in the EU´s energy policy, setting up an overall binding EU target[2] of 20% RES in gross final energy consumption by 2020, in addition to binding national targets ranging between 10% (Malta) and 49% (Sweden) that reflect the various circumstances – e.g., differences in renewable energy potential and economic performance- and starting points in each country. The same Directive also requires the Member States to develop national action plans (so-called National Renewable Energy Action Plans -NREAPs) to stablish a roadmap for development of RES and the creation of cooperation mechanisms between states to achieve the target in accordance with the cost-effectiveness principle.

Along the same line, the EU Directive 2018/2001 on the promotion of the use of energy from renewable sources (RED II) established a framework for meeting a binding EU target of at least 32% renewable energy in gross final energy consumption by 2030. More recently, the goal of reaching climate neutrality (through green technology, sustainable industry and transport, and cutting pollution) by the horizon of year 2050 has been embedded at the very core of the European Green Deal (European Commission, COM2019).

The role of renewable energies is also emphasized in the Energy System Integration Strategy (COM2020), and more recently in the context of the Recovery and Resilience Facility (EU Commission; May 2020), where clean and renewable energy sources will play an important part in the recovery plans in the aftermath of the Covid-19 crisis.

The above Directives have been supported by significant investments in order to ensure their objectives. Cohesion policy instruments –more specifically, the European Regional Development and Cohesion Funds- are the most important funding sources within the EU spending programs to promote renewable energy, with investments in sustainable energy (including energy efficiency, renewable energy, smart energy infrastructures, and low-carbon research and innovation) totaling around €30 billion.

---

[2] A 20.6% overall target for the EU-27.

While sharing a common policy and targets concerning renewables, the group of countries included in the EU is relatively diverse, with different antecedents and idiosyncratic features deriving from economic, social, cultural, and climate differences, so our primary goal in this paper will be the analysis of convergence in the RES share indicators of the EU-28.

Convergence in environmental variables is relevant because of its policy implications, as it is expected that those countries that converge will be able to more effectively implement common environmental policies (e.g., Aldy, 2006; Herrerias, 2013; Burnett, 2016; Apergis *et al.*, 2017) such as those devised by the EU. Furthermore, convergence would also imply that countries are converging toward a common understanding of global threats, as well as being potentially helpful to determine the efficiency and speed of environmental policies (Bilgili and Ulucak, 2018).

The interest in the analysis of convergence in environmental variables primarily stems from the links between growth and environment, which may be modelled through the Environmental Kuznet's Curve (EKC) (Grossman and Krueger, 1991). As pointed out by Brock and Taylor (2004), a proper understanding of that relationship may be a key driver to long run prosperity. In this regard, Strazicich and List (2003) was among the pioneering works to explore the connections between the empirical research line that correlates pollution with income and the literature on convergence of spatial incomes over time. More recently, Stern (2017) --by combining the EKC and some convergence approaches—has also shown the relevance of convergence for explaining both pollution emissions and concentrations. The interest in convergence is also motivated by the global mitigation efforts of both governments and international organizations to stop climate change and global warming, which has become one of the main challenges for this century.

Many previous studies have addressed convergence analysis for various environmental magnitudes, most of them focusing on issues like carbon dioxide and greenhouse gas emissions, ecological footprint, or energy productivity and consumption. This paper has as its distinctive features that convergence is analyzed from the standpoint of renewables (which have been less frequently studied in convergence studies and are a key factor to reduce emissions and bring the EU closer to its neutrality goal) and that the study is conducted at the EU level.

We shall rely for the analysis on the club convergence approach as proposed by Phillips and Sul (2007, 2009). In addition to testing for convergence, this framework also enables a systematic analysis of club convergence issues, including an algorithm that allows detection of groups of EU members that have shifted from disequilibrium to specific steady-state positions.

More precisely, this paper focuses on analyzing the potential presence of clubs of convergence in the RES share indicator (defined as the share of renewable energy in gross final energy consumption). The problem may be addressed from two alternative standpoints, either *global convergence* to the common objective of 20% or *specific convergence* of each Member State to its target. The analysis will be complemented with the results of a probit model to help clarify some potential determinants for the cluster structures detected. Finally, with the aim of studying convergence at the sectoral level, we also analyze potential clubs of convergence in the electricity, transport, and heating and cooling sectors.

The remainder of the paper is organized as follows: Section 2 includes a review of literature. Section 3 briefly describes the methodology of club convergence analysis and the dataset employed for the analysis. Our main results and a discussion are included in Section 4. Finally, Section 5 outlines some conclusions.

## 2. Literature review

Following the seminal paper by Barro and Sala-i-Martin (1990), convergence analysis has attracted a great deal of attention in many areas of macroeconomic theory and has extended to many other fields including the study of environmental variables, pioneered by List (1999) and Strazicich and List (2003). At the same time, empirical analyses have been relying on various methodologies —-including time series, panel, and cross-section studies-- and have studied several (including beta, sigma, club, and stochastic) convergence concepts[3]. In this paper we analyze a data panel, with a focus in detecting the potential presence of convergence clubs, defined as groups of economies with similar conditions and structural characteristics (such as technology, preferences, and political systems) converging to the same steady state.

One of the earliest applications of the club-convergence approach to environmental variables is Panopoulou and Pantelidis (2009). Thereafter, this technique has been employed for the analysis of various environmental variables. These include:

(i) *Carbon dioxide emissions for specific geographical areas:* This research line includes, among others, Panopoulou and Pantelidis (2009), who analyze a large database of 128 countries, Herrerias (2013) who conducts a study for a large group of both developed and developing nations, Wang *et al.* (2014), for Chinese provinces, and Burnett (2016) and Apergis and Payne (2017), for U.S. states. More recently, Morales-Lage *et al.* (2019) study per capita $CO_2$ emissions in EU-28 countries, by energy subsectors, and Haider and Akram (2019a) analyze per capita carbon dioxide emissions and their components (coal, oil, and gas) for a group of 53 nations.

(ii) *Greenhouse gas emissions:* e.g., Ivanovski and Churchill (2020) for Australian regions and Presno *et al.* (2021) for EU-28 states.

(iii) *Ecological footprint:* Recent works that exploit this increasingly employed multidimensional indicator include Ulucak and Apergis (2018), for EU members; Haider and Akram (2019b), who analyze convergence in both per capita ecological and carbon footprints for a sample of 77 countries; and Solarin *et al.* (2019), for a panel of 92 nations. The works by Erdogan and Okumus (2021) and Bilgili and Ulucak (2019) also rely on the ecological footprint paradigm, combining club convergence and panel unit-root/stationarity testing.

(iv) *Carbon intensity:* e.g., Emir *et al.* (2019), for EU-28 countries, Camarero *et al.* (2013), for OECD countries, and Yu *et al.* (2018), for Chinese industrial sectors.

---

[3] Other studies in the field of convergence of environmental variables (e.g., among others, Strazicich and List, 2003; Presno *et al.*, 2018; Churchill *et al.*, 2020) have analyzed stochastic convergence by employing unit root testing methods. See Ulucak and Apergis (2018) for an extensive review of the environmental convergence literature.

(v) *Energy productivity*: This indicator has been extensively analyzed. Two recent papers in this area are due to Bhattacharya *et al.* (2018; 2020), who study, respectively, panels of Indian and Australian states and territories. A thorough review of the literature on energy productivity convergence appears in Bhattacharya *et al.* (2018).

(vi) *Energy consumption:* Ivanovski *et al*. (2018) analyze convergence in per capita energy consumption at sectoral level (with nine sectors being considered) across Australian regions. Herrerias *et al.* (2017) apply the same approach to energy consumption[4] in the residential sector across Chinese regions and Kim (2015) studies convergence in per capita electricity consumption for a set of 109 countries.

Some recent papers have specifically focused on convergence in renewable energy consumption. Among others, Reboredo (2015), Solarin *et at*. (2018), and Demir and Cergibozan (2020), who analyzed panels of 39, 27, and 28 OECD countries, respectively, and Payne *et al*. (2017), who conducted the analysis for the 50 states of the US. Studies devoted to EU countries have been less frequent. In this regard, Berk *et al.* (2020) applies a System Generalized Method of Moments methodology to the EU-14 group of countries, analyzing the contribution of renewable energy sources to primary energy consumption and finding evidence of both absolute and conditional convergence. Kasman and Kasman (2020) also address convergence in per capita renewable energy consumption across EU-15 member states, from both the stochastic convergence and the club convergence standpoints, detecting the presence of three clubs plus a divergent country (Denmark). Also for EU states, and using both parametric and semi-parametric methods, Butnaru *et al.* (2020) conclude that there exists a relationship between conventional and renewable per capita energy consumption convergence.

---

[4] See Herrerias *et al.* (2017) for a review of papers analyzing convergence in energy consumption through several methodologies.

## 3. Methodology and data

### 3.1. The club convergence approach

The following single factor model is assumed:

$$y_{it} = \delta_{it}\mu_t, \text{ for all } i, t \tag{1}$$

where $y_{it}$ denotes the RES share indicator for country $i = 1, …, N$ and year $t = 1, …, T$; $\delta_{it}$ is the "transition parameter", that is, a time-varying idiosyncratic element; $\mu_t$ denotes a "common growth component" that aims at capturing some deterministic/stochastic trending behavior in $y_{it}$.

Phillips and Sul (2007) define the following relative transition parameter:

$$h_{it} = \frac{y_{it}}{N^{-1}\sum_{i=1}^{N} y_{it}} = \frac{\delta_{it}}{N^{-1}\sum_{i=1}^{N} \delta_{it}} \tag{2}$$

which traces out the individual transition path over time for economy $i$ in relation to the panel average. Under convergence, $h_{it}$ converges in probability to 1 for all $i$ as $t \to \infty$, whereas the cross-sectional variance of $h_{it}$, namely $H_t = N^{-1}\sum_{i=1}^{N}(h_{it} - 1)^2$, converges to zero.

Philips and Sul (2007) proposed the *logt* convergence test, which involves estimating by OLS (with HAC standard error estimates) the following equation:

$$\log\left(\frac{H_1}{H_t}\right) - 2\log(\log(t)) = a + b\log(t) + \varepsilon_t \tag{3}$$

with $t = [rT], [rT] + 1, …, T$, with $[rT]$ being the integer part of $rT$, and the value $r = 0.3$ suggested for small/moderate sample sizes ($T \leq 50$).

Under the null of convergence, the least squares estimator of $b$ converges in probability to the scaled speed of convergence parameter ($2\alpha$), and the null hypothesis can be tested through a one-sided $t$ test of $b \geq 0$, with rejection at 5% significance when $t_{\hat{b}} < -1.65$. Given the relationship $b = 2\alpha$, the case $b \geq 2$ (or equivalently, $\alpha \geq 1$) means absolute convergence (i.e., convergence in levels), whereas $0 \leq b < 2$ implies conditional convergence (i.e., the growth rates converge over time).

When the null hypothesis of convergence for the whole panel is rejected, Phillips and Sul (2007) propose a clustering algorithm, based on the application of the *logt* test, which allows the endogenous identification of specific subgroups that converge (so-called clubs of convergence).

The *logt* convergence test can also be employed to merge the initial clubs into larger groups, with the aim of both correcting for the tendency of the algorithm to over-estimate the number of clubs (Phillips and Sul, 2009) and investigating so-called *transitions* between clubs (namely components of a club moving to a contiguous group, or clubs slowly converging to one another).

## 3.2. Data

With a view to assessing the progress towards its RES objectives, the EU computes[5] the overall RES share indicator. The numerator of that ratio is the renewable[6] energy delivered to final consumers (industry, transport, households, services, agriculture, forestry, and fisheries); the denominator is the gross final energy consumption of all energy sources, that is, the energy employed by end-consumers (final energy consumption) plus grid losses and self-consumption of power plants. The source of data is the European Environment Agency (via Eurostat).

The RES indicator assesses how extensive the use of renewable energy is and, indirectly, the degree to which renewable fuels have replaced fossil and nuclear fuels and therefore contributed to decarbonization of the economy. The RES share is part of several indicator sets. More specifically, it is included among the EU 2020 strategy indicators, monitoring the progress towards three of the EU Sustainable Development Goals (SDG), namely SDG 7 (on affordable and clean energy), SDG 12 (on ensuring sustainable consumption and production patterns), and SDG 13 (on climate action). It is also integrated in the impact indicators for the Strategic Plan 2016-2020 referring to the 10 Commission priorities, and in the set of indicators for the EU´s framework Strategy for a Resilient Energy Union.

---

[5] The indicator is computed according to the Renewable Energy Directive (Directive 2009/28/EC) on the promotion of the use of energy from renewable sources. It is calculated on data collected under Regulation 1099/2008 (EC) on energy statistics, complemented by specific supplementary data transmitted by the national administrations to Eurostat. More details on the calculation methodology applied by Eurostat can be found in "SHARES Tool Manual". (EUROPEAN COMMISSION EUROSTAT Directorate E: Sectoral and regional statistics Unit E.5: Energy). https://ec.europa.eu/eurostat/documents/38154/4956088/SHARES+tool+manual-2019.pdf/8822f775-121e-f73e-3040-93c01f6ebd3e

[6] Renewable energy can be generated from a wide range of resources: sun (concentrated/photovoltaic power), hydro (large, small, and micro hydro, including tide, wave and ocean energy), wind (onshore/ offshore), geothermal energy and all forms of biomass (including solid, liquid, landfill gas and biodegradable fraction of waste, and liquid biofuels). The contribution of renewable energy from heat pumps is covered for those Member States for which that information is reported. Exports/imports of electricity are not considered in this indicator, although statistical transfers and other flexibility measures reported to Eurostat and complying with the requirements of Articles 6-11 of Directive 2009/28/EC are included. Currently, only Sweden (with Norway) and Luxembourg (with Estonia and Lithuania) are resorting to those flexibility measures.

In addition to the overall RES share indicator, Eurostat also publishes the share of renewable energy disaggregated by the following three sectors:

1) Transport (RES-T share), computed as "energy from renewable sources consumed in transport divided by the total amount of energy consumed in transport"[7].

2) Heating and cooling (RES-H&C share), defined as "gross final consumption of energy from renewable sources for heating and cooling divided by gross final consumption of energy for heating and cooling".

3) Electricity (RES-E share), which is calculated as "gross final consumption of electricity from renewable sources divided by gross final consumption of electricity".

Our study focuses on the 2004-2018 period, with the raw series (instead of their logarithms)[8] being employed for the convergence analysis.

**4. Results and Discussion**

According to the RES share indicator, the use of renewable energy has been increasing continuously in the EU along the 2004-2018 period, with its share more than doubling since 2004, when renewables covered only 8.5% of gross final energy consumption. In year 2018, the EU reached a share of 18% (18.9% in EU-27), which is above the 16% indicative target for 2017/2018.

Disaggregated by member states, wide variations are observed in 2018 in the shares of renewable energy on final energy consumption: Sweden led with 54.7%, followed by Finland and Latvia with shares slightly above 40%; in the other end we find Malta and the Netherlands, with shares below 8%. When compared with their respective targets, twelve countries exhibit renewable energy shares above their 2020 targets, and eleven met or exceeded their RED I average indicative trajectory for 2017-2018, with the exceptions being France, Ireland, the Netherlands, Poland, and Slovenia. These shares reflect the historic heterogeneity in EU States´ energy mix and their differences in terms of economic development, renewable energy potentials, and financial and regulatory support.

---

[7] Article 3(4) of the RES Directive (2009/28/EC) specifies that "only petrol, diesel, biofuels consumed in road and rail transport, and electricity, including electricity used for the production of renewable liquid and gaseous transport fuels of non-biological origin, shall be taken into account". Biofuels would also include liquid or gaseous fuel for transport that is produced from biomass.

[8] This prevents information loss when eliminating those countries having null values in their RES magnitudes on the first years of the study period.

The share of renewables increased in the three application areas. In year 2018, overall for the EU, the share of renewables was highest in electricity generation (32%), followed by heating and cooling (20%) and transport (only 8%). However, in terms of consumption of renewable energy, heating and cooling was the biggest contributor in that year, followed by electricity and transport. At EU level, RES-E and RES-H&C are above the targets defined by states in their NREAPs, while the transport sector is slightly below, possibly as consequence of some uncertainty on biofuels policy and the adjustments it entails in the legislative framework.

### 4.1. Overall RES share

We first carry out a convergence club analysis for the overall RES share. Then, with a view to gaining further insight on the determinants of the cluster classification delivered by the Phillips and Sul (2007) algorithm, a probit model will be fitted to data.

### 4.1.1. Club-of-convergence results

Global convergence may be studied either directly for the RES series or by analysing the rescaled series, after dividing the observed RES values by the 20% target. Both approaches are equivalent as the later involves only a change of scale in the indicator, with the invariance properties of the *logt* test and club convergence algorithms implying that the results are identical in both cases.

[PLEASE INSERT TABLE 1 ABOUT HERE]

The null of overall (*global*) convergence is rejected (see Table 1 below). The algorithm detects two convergence clubs. Club 1 brings together the "top-performing" countries (i.e., those having national 2020 RES targets over 30%), namely Sweden, Finland, Latvia[9], Denmark, Austria, and Portugal. On the contrary, Club 2 includes states with RES targets not exceeding 25%.

The transition path of Club 1 is uniformly above unity (see Figure 1) and doubles the average of the UE, mainly because of the high RES values of Sweden, Finland, and Latvia. However, a decrease in the transition path of that club is observed in the most recent years, along with a slow and less pronounced rise in the transition path of Club 2. In addition, conditional convergence is found in both clubs, with a very low speed of convergence (see the α values in Table 1).

[PLEASE INSERT FIGURE 1 ABOUT HERE]

With a view to more closely examine both clubs, Figure 2 below displays the transition paths of each country in relation to the mean of its club. As expected under club convergence, a gradual reduction in dispersion is observed. It is remarkable the transition path corresponding to Sweden (see Figure 2.a), whose values are far away from the rest of countries in Club 1. The positive evolution -relative to the mean of the club- of Denmark is also noteworthy. The reverse holds true for Latvia, and to a lesser extent for Finland. Figure 2.b shows the performance of the members of Club 2. A positive evolution (relative to the mean of the club) is observed for the United Kingdom, Belgium, Luxembourg, and Malta. The reverse occurs for Estonia, Croatia, Lithuania, Romania, and Slovenia.

[PLEASE INSERT FIGURE 2 ABOUT HERE]

The small number of clubs identified by the algorithm contrasts with the results obtained when the shorter series are analyzed. Thus, when only the 2004-2016 period is considered, and employing the merging procedure, three clubs plus a divergent country (Sweden) are detected. This may be taken as an indication that the process of convergence has been going on throughout the study period. With this in mind, we also checked for potential transitions between clubs, but the results of the algorithm led us to reject that possibility (see Table 1).

---

[9] Latvia is ahead of its indicative RED and planned NREAP trajectory for 2015-2016, although this has been a consequence of lower energy consumption, rather than of reaching the levels of gross RES consumption as planned.

As commented in previous sections, given the wide diversity of national RES goals, it is also interesting to conduct the same analysis from the standpoint of *specific* convergence of each country to its own RES target. In this case the variable of interest is the ratio of country $i$'s RES value to country $i$'s RES target, which is readily interpreted as the 'distance' of country $i$ ($i = 1, ... , 28$) to its specific RES goal for year 2020. When the *logt* convergence test is applied (see Table 1, panel B) the null of convergence is not rejected, suggesting that the renewable energy efforts of the UE members are at least converging to their respective national targets as assigned by the UE Directive 2009/28/EC, although possibly not to a single fixed steady state.

### 4.1.2. A probit model for club membership

With a view to gaining further insight on the above results, a binary probit model was fitted. The variable to predict was club membership (equal to 0 for the countries in Club 1 and to 1 for those in Club 2). The small sample size (only 28 cases) available severely limited model complexity. The explicative variables considered appear in Table 2 below and were obtained from Eurostat. Their values are averages for each country over the 2009-2018 period (a shorter range had to be considered in some cases due to limitations in data availability).

[PLEASE INSERT TABLE 2 ABOUT HERE]

Real GDP per capita (GDPCAP) is usually considered in literature as a relevant, positive indicator of renewable energy diffusion. Many papers (among others, Carley, 2009; Sadorsky, 2009, a-b; Burke, 2010; Gan and Smith, 2011; Pfeiffer and Mulder, 2013; Apergis and Payne, 2014a-b; Kim and Park, 2016; da Silva *et al.*, 2018) have detected, for several geographical contexts and employing various methodologies, statistically significant evidence of a direct relationship between income (e.g., real GDP per capita) and renewable energy use (either per capita or in terms of its share in overall energy consumption). A recent work by Ohler (2015) has also found empirical evidence (for the case of the US electricity mix) of a nonlinear, U-shaped relationship between RES and GDP per capita that can be justified in terms of empirical Kuznets curve theories. These considerations led us to include both GDPCAP and its square in the probit model.

The ratio of national expenditure on environmental protection to nominal GDP (ENVEXPGDP) tries to summarize the relative environmental effort carried out by each country along the study period. It may be connected (in an imperfect way) with many qualitative factors--including policies promoting renewables and continuous commitment to renewables--which have been found (e.g., Marques and Fuinhas, 2012; Aguirre and Ibikunle, 2014) to a have a positive link to renewables diffusion.

Energy import dependency (ENIMPDEP) indicates the average dependency from energy imports along the study period. Some studies (e.g., Marques and Fuinhas, 2012) report a statistically significant, inverse relationship between energy import dependence and renewable energy consumption.

The inclusion of nuclear enrichment capacity (NUCLENRCAP) obeys a somewhat different rationale, as nuclear appears as a relevant energy source in several countries that might be employing it as a complementary tool to reduce carbon emissions.

[PLEASE INSERT TABLE 3 ABOUT HERE]

Results in Table 3 above indicate that the model has a suitable performance in terms of goodness of fit diagnostics, with almost 93% of the cases correctly classified and all the explicative variables being statistically significant.[10]

GDPCAP and its square are both statistically significant (though only marginally in the latter case), with SQ_GDPCAP having a positive coefficient that would be in accordance with the potential presence of a nonlinear (U-shaped) effect of GDP per capita on the probability of belonging to Club 2. This may be reasonable as Club 2 is quite heterogeneous, including both most Eastern and Southern countries (which have relatively low GDP per capita) and some economies (like Luxembourg, the Netherlands, and more recently Ireland) with very high GDP per capita values.

---

[10] Admittedly, the above probit model may be somewhat naïve, but sample size greatly limits the possibility of fitting more complex structures and including additional explicative variables. Some variables that were highly correlated with those in Table 2 above also had to be discarded in advance, as well as others that essentially were mere redefinitions of the RES indicator or were related to it by more or less direct accounting identities.

As for ENVEXPGDP, the model estimates a negative (strongly significant) marginal effect on the probability of being included in Club 2. This also seems expectable as Club 1 includes some countries (particularly Austria) where environmental expenditure had a remarkably high share in GDP on average in the study period, as compared with most EU members.

ENIMDEP also has a strongly significant, positive coefficient, indicating that high dependence on foreign energy imports tends to increase the probability of being in Club 2, all the other things being equal. This can also be expected as Club 1 includes countries (like Denmark and Sweden) having a low dependency on imported energy, whereas Club 2 clusters economies (remarkably, Malta, Cyprus, Luxembourg, Ireland, Portugal, and Spain) having the highest dependency levels.

Finally, NUCLENCAP seems the most statistically significant variable in the model, also having a positive effect on the probability of being included in Club 2. This is also far from surprising as no member of Club 1 has nuclear capacity, which concentrates in France, Germany, and the Netherlands, all of them in Club 2.

**4.2. Clubs of convergence in specific sectors**

In this subsection we analyze convergence clubs for the three sectors where disaggregated data are available.

**4.2.1. RES-Transport**

Since 2004 the share of renewables in transport has increased fivefold, up from only 1.4%. Renewables consumed in transport in the EU mostly come from liquid biofuels[11] (estimates suggest that the share of biofuels in the RES-T indicator is around 89%), although Member States are increasingly promoting e-mobility options, through the implementation of subsidies.

A 10% RES-T target[12] has been stablished by 2020 at EU level, which translates into the same goal for all Member States. However, only two countries (Sweden, 29.7%; Finland, 14.9%) exceeded that target in 2018, and four others (the Netherlands, Austria, France, and Portugal) were in the 9%-10% range. For the remaining 22 Member States, shares were between 2.7% (Cyprus) and 8% (Bulgaria). Additionally, transport is the only sector whose level is below the target defined by the Member States in their NREAPs (8.03% -for the EU28- versus 8.5% planned).

From a temporal perspective, following rapid growth between 2005 and 2010, RES-T dropped in 2011 and has been increasing at a slower pace since 2012, with high volatility being observed in some countries. The most remarkable case is Finland, whose RES-T share decreased by 16 percentage points between 2015 and 2016, increasing again by 10 percentage points in 2017 as consequence of a drop in national use of biodiesel, in parallel with higher exports in 2016. A less dramatic case is Spain, whose indicator grew by 4 percentage points between 2015 and 2016 because of the implementation of a new information system to register biofuel certifications. 2011 was another year when some countries (e.g., Bulgaria, Portugal, France, Finland, Spain) experienced abrupt changes in their indicators. Evidently, this has been reflected in the overall RES-T indicator and -as shown in Figure 3 below- also in the shapes of the transition paths.

[PLEASE INSERT FIGURE 3 ABOUT HERE]

The null hypothesis of convergence in RES-T is rejected, with the following two clubs being detected:

---

[11] Most biofuels consumed in the EU are biodiesel, with around 59% of the feedstock for biodiesel consumed in year 2018 being either imported or produced from imported feedstock. In that year, the total amount of cropland dedicated to biofuel production in the UE was 3%. Some studies (e.g., Filip *et al*., 2019) have found a very weak evidence of biofuels affecting food prices in times of scarcity.

[12] Only biofuels complying with the sustainability criteria of the EU Renewable Energy Directive are to be counted towards this target.

- Club 1 is integrated by the two countries (Sweden and Finland) that widely exceed the 10% objective, plus Bulgaria, Malta, Denmark, and Romania. The inclusion of Malta in Club 1 may be due to its great progress in recent years, taking off from 0% in year 2010 to almost 8% in 2018. The transition path in that club is swinging, mainly as a consequence of the above-mentioned erratic behavior of the indicator in Finland. On the contrary, Sweden exhibited a distinctively rising evolution throughout the period, almost increasing its share fivefold, mainly as consequence of the use of biofuels.
- Club 2 is a heterogeneous group that includes the rest of the EU28 countries.

**4.2.2. RES-Heating and Cooling**

Heating and Cooling in buildings and industry accounts for half of the energy consumed in the EU. Around 20% of heating and cooling is generated from renewable energy (with solid biomass being the main source, followed by heat pumps, biogas, and solar thermal collectors), which has a crucial role and an enormous potential in the EU´s energy and climate policies. Indeed, it is projected that about 40% of the 32% target share of renewable energy by 2030 will come from the H&C sectors.

During the 2004-2018 period, the RES-H&C value almost doubled in the EU; however, when the individual Member States are examined, considerable differences among them are observed, with the indicator ranging from 65.4% (Sweden) to 6.1% (the Netherlands) in 2018.

The null of convergence is rejected for the H&C sector, with three clubs identified: Club 1 is led by a large portion of members of the Council of the Baltic Sea States (Sweden, Latvia, Finland, Estonia, Denmark, and Lithuania) plus two islands (Cyprus and Malta). Both Club 1 and 2 (with the latter including Portugal, Croatia, Austria, Bulgaria, Slovenia, Greece, Romania, France, and Czechia) display paths over 1 (see Figure 4), that is, above the mean of the EU. The null hypothesis of these two clubs being merged is rejected, but not that of a transition between them (see Table 1). The rest of EU countries are classified in Club 3, with their transition paths far away from 1.

[PLEASE INSERT FIGURE 4 ABOUT HERE]

### 4.2.3. RES-Electricity

The electricity sector has experienced a take-off in renewables in recent years, pushed up by a sharp growth in both solar photovoltaics and wind (fostered by the decline of electricity costs from these two sources) and a significant contribution of hydropower generation, with the result that RES-E more than doubled.

The analysis of convergence allows us to conclude that electricity is the only sector for which Members States achieved convergence in renewables. Despite the initial diversity (e.g., in year 2004, Austria and Sweden´s shares were 61.6% and 51.2%, respectively, whereas Malta and Cyprus had 0%), dispersion has been decreasing along the study period, as revealed by the transition paths (see Figure 5 above), although the speed of transition seems to be low ($\hat{\alpha}$ =0.027).

[PLEASE INSERT FIGURE 5 ABOUT HERE]

### 4.3. Discussion

When comparing overall and sector results it is remarkable that Sweden, Finland, and Denmark are in all cases positioned in clubs that outperform the mean of the EU. This seems far from for surprising since the policies in those three countries have been especially proactive on the environmental front. Latvia, also a member of the Council of the Baltic Sea States, is included in Club 1 because of its favorable evolution in the electricity and H&C sectors, although the country has had a more lackluster progress in transport.

Two seemingly different countries (Austria and Portugal) are nevertheless classified in the same group (Club 2, although transitioning towards Club 1) in both the transport and H&C sectors, also sharing top places in electricity (where Austria has been the EU leader throughout the period), all of which positions both states in Club 1 in terms of overall RES.

Germany, France, and the United Kingdom (the largest three Member States, with a joint share in gross inland consumption that approaches 50% of EU-28) do not play leading roles in any of the "best positioned" clubs. Interestingly, they all are either located into the same clubs or classified as transitioning (e.g., in the H&C sector). Despite having a higher RES share than the other two countries, France is at severe risk of failing to meet its target. The same holds true for the Netherlands, which (excepting the transport sector) is at the bottom of the worst positioned clubs and has agreed with Denmark (under the co-operation mechanisms stipulated in RED I) a statistical transfer to facilitate the achievement of its target. An analogous formula has been employed by Luxembourg and Malta.

The analysis disaggregated by sector indicates that only electricity exhibits convergence. As for Heating and Cooling, it has a large potential in the EU´s energy and climate policies, in the light of its high percentage in the energy consumed and the relatively small share of heating and cooling generated from renewable sources. An enormous potential can also be found in the transport sector. This was the smallest contributor in year 2018 (in terms of consumption of renewable energy) and remains the only sector whose share of renewables is below the target defined by the Member States in their NREAPs. That lag may be partly attributed to relatively high abatement costs related to biofuels and the uncertainty on the policy framework, that slows down progress in the deployment of second-generation biofuels. In this regard, the EU has decided to minimize the use of food and feed-crop-based biofuels (a 7% cap on the portion of biofuels made from crops has been proposed, in order to avoid deforestation and stress on land resources) and to focus in the future on promoting advanced biofuels and other low carbon fuels, such as renewable electricity, recycled carbon fuels, and renewable liquid and gaseous transport fuels of non-biological origin. As commented above, the Phillips-Sul algorithm finds two clubs in this sector. The first one is headed by Sweden and Finland, with both countries far exceeding the 10% RES-T target. Some Member States classified in Club 2 are in severe risk of not fulfilling that legal obligation, so they should take appropriate actions, either by national deployment or via cooperation mechanics like statistical transfers[13] for the transport sector, as enabled under the Indirect Land-use Change –ILUC- Directive.

## 5. Conclusions

In this paper we have analyzed convergence in the RES share for the countries in the EU-28. The study period includes the most recent data available (up to year 2018, two years away from the 2020 objective's deadline). The EU-28 is a diverse group of countries in terms of both their starting points and historic antecedents concerning the use of renewables and their economic, cultural, climatic, and social characteristics, although with the common overall objective of attaining a 20% target of renewable energy in gross final energy consumption by 2020, as a first step towards more ambitious goals of at least 32% in 2030, and climate-neutrality (i.e., an economy with net zero greenhouse gas emissions) by the horizon of year 2050.

Convergence analysis is relevant to examine both the speed and efficiency of environmental policies. Here we have relied on a club-of-convergence approach that makes it possible both to obtain groups of countries that evolve from disequilibrium to specific steady-state positions and to estimate their speed of convergence. Subsequently, those groups of countries could adopt common environmental policies.

The study was conducted from two complementary standpoints: global convergence to the common 20% objective and specific convergence of each Member State to its objective. In the latter case, so-called conditional convergence was detected (meaning that only the growth rates, but not the national levels of the RES indicators, converge over time), although the estimated convergence rate is slow.

A somewhat different (but not necessarily incompatible) picture emerges when the convergence analysis is carried out with respect to the overall-EU 20% objective: two clubs of convergence are detected, with the first one corresponding to states with national 2020 RES targets over 30% and the second club including those countries with targets below 25%. Conditional convergence, at very slow rates, is observed in both clubs. This result has a primary implication: our finding of two clubs may indicate the convenience of devising more specific policies as well as regulations oriented to reduce heterogeneity. Countries located in Club 2 would demand more effective public initiatives to incentivize fulfillment of the overall goal, in addition to an accelerated implementation of new policies. Different countries would require different, specifically adapted measures and targets to be implemented. Regarding this, the probit analysis that complements the study may provide some helpful guidelines on those policies, as it suggests that club membership might be explained by a simple model including variables like real GDP per capita (having a U-shaped effect), relative economic effort in environmental protection, dependency on energy imports, and nuclear capacity. All those variables would be statistically significant, with positive coefficients obtained for the latter two (indicating that, all the other things being equal, high values of those variables tend to rise the probability of a country being in Club 2, which groups the worst performing countries). The reverse would occur for the remaining variables in the model. In the case of nuclear capacity, this might be in accordance with the possibility that some countries may be taking advantage of that energy source as a help to reduce GHG emissions. The U-shaped effect detected for real GDP per capita may possibly agree with the potential presence of a nonlinear, Kuznets-curve-type marginal effect of that variable, that might affect club membership (itself closely related to each country's RES goal). This evidence would agree with recent findings by Ohler (2015). Overall, the results of probit modelling may suggest that policies should be oriented to increasing the efforts in environmental protection and reducing dependency. Regarding this, some useful guidelines for the EU case may also be found in Marques and Fuinhas (2012), where several public policies that have been major drivers in the development of renewables are enumerated, with the role of incentives/subsidies (including feed-in tariffs) and "policy processes" (such as strategic planning) being emphasized. As for dependency, the same study concludes that it is mainly concentrated in traditional energy sources, indicating the presence of productive infra-structures that heavily rely on fossil fuels and may pose considerable barriers to a transition to renewables. While it is true that in the past the deployment of renewables has mostly policy-driven, it seems reasonable to enhance the role of market logic in this process. Consumers are increasingly conscious on the need for climate action and the environmental effects of their personal economic choices. This gradual shift to environmental sustainability in the demand of goods and services should be naturally accompanied by adaptions on the supply side with a view to fulfilling those needs, so it seems reasonable to expect that –although relying on different methods and incentive schemes--public policies and market action will naturally evolve along quite close, environmentally conscious pathways. EU citizens, in their double political/economic role, will be clearly pushing up that approach.

---

[13] Member States are also allowed to fill their yearly quotas in advance through statistical transfers (both between years and to other states) and variation in blends between gasoline and diesel.

The study was complemented with an analysis disaggregated by sector, which may also suggest additional policy implications. Results indicate that only electricity would exhibit convergence, possibly as a consequence of the vigorous growth in solar photovoltaics and wind (as a result of the drop in electricity costs from these two sources), supplemented by hydropower generation. In any event, in addition to continuing to innovate in the electricity sector, there is still a long way to go, with both further progress in deployment of renewables for heating and cooling and extended use of renewable transport options clearly on the roadmap.

In the case of the transport sector, market uncertainties arising from policy changes in the field of biofuels have been unfavorable, so it is important to design a policy framework that provides a safe and stable legal framework to all stakeholders. Also, some serious difficulties in this sector may stem from the lack of appropriate infrastructure for electric vehicles and alternative fuels, so investment in that area will be much needed. It is also important to intensify efforts to promote advanced biofuels, renewable fuels of non-biological origin, and recycled and low carbon fuels, by stimulating the use of liquid biofuels both in the current stock of vehicles and in the transport modes where electrification is currently unfeasible, also incentivizing the use of electric and hybrid vehicles.

Certainly, several plans like the European Strategy for Low-Emission Mobility (which strongly focuses on road transport), the Seventh Environment Action Programme, the EU plans on Accelerating Clean Energy Innovation (which set up among their overarching goals positioning Europe as the global leader in renewables), and the European Green Deal, constitute relevant milestones on the way to increasing use of renewables in the transport subsector, although these efforts must be sustained in the future. In this regard, the Recovery and Resilience Facility aims at providing the investments required for the green and digital transitions, as well as encouraging Member States to put forward investment and reform plans in certain areas, including clean technologies and renewables, sustainable transport, and charging and refueling stations.

Barriers also remain in the heating and cooling sector, mainly from deficiencies in the capacities of district heating networks, although the EU is carrying out active investment and research in efficient systems in that area.

In the case of the electricity sector, both spatial planning needs and environmental regulations have been limiting the advances in some Member States. Integrating the increasing RES capabilities into the grid remains a challenge for most countries, due to the high costs of grid connecting and the uncertainties on the various scenarios of grid development and transparency in the connection procedures.

The EU has set up as one of its priorities to become a world leader in renewable energies, that should have a presence in all dimensions of the Energy Union. For that purpose, technology research and large investments have been done, although those same efforts should also yield substantial benefits in areas including employment, growth, trade balance, and industrial competitiveness, stimulating the emergence of a new industrial base which has a major source in the renewables sector. Certainly, achieving technological leadership is central for the clean energy sector, but a permanent effort is also required to achieve competitive advantages in many areas, like batteries.

Finally, the current Covid-19 pandemic is still having a significant impact on energy demand. A favorable, side effect of this exceptional situation has been that the renewable energy shares projected for year 2020 have generally risen, although some caution is required when interpreting those advances as it remains possible that some of those increases are temporary and may eventually be reduced as economic activity reverts to previous levels. In this regard, the financial injection to be provided by EU´s 2021-2027 long-term budget, supplemented by so-called *Next Generation EU* recovery instrument (especially in its Headings 1 and 3), also represents a serious effort to repair the damage caused by the coronavirus pandemic, as well as an opportunity to strongly accelerate the transition towards a more sustainable Europe.

# REFERENCES

Aguirre, M., Ibikunle, G., 2014. Determinants of renewable energy growth: A global sample analysis. Energy Policy 69, 374-384.

Aldy, J.E., 2006. Per capita carbon dioxide emissions: convergence or divergence? Environmental and Resource Economics 33, 533-555.

Apergis, N., Payne, J.E., 2014a. The causal dynamics between renewable energy, real GDP, emissions and oil prices: evidence from OECD countries. Applied Economics 46(36), 4519-4525.

Apergis, N., Payne, J.E., 2014b. Renewable energy, output, CO2 emissions, and fossil fuel prices in Central America: Evidence from a nonlinear panel smooth transition vector error correction model. Energy Economics 42, 226-232.

Apergis, N., Payne, J.E., 2017. Per capita carbon dioxide emissions across U.S. states by sector and fossil fuel source: Evidence from club convergence tests. Energy Economics 63, 365-372.

Apergis, N., Payne, J.E., Topcu, M., 2017. Some empirics on the convergence of carbon dioxide emissions intensity across US states. Energy Sources Part B: Economics, Planning and Policy 12(9), 831-837.

Barro, R.J., Sala-i-Martin, X. 1990. Economic growth and convergence across the United States. National Bureau of Economic Research. Working Paper W3419.

Berk, I., Kasman, A., Kılınç, D., 2018. Towards a common renewable future: The System-GMM approach to assess the convergence in renewable energy consumption of EU countries. Energy Economics 87, 103922.

Bhattacharya, M., Inekwe, J.N., Sadorsky, P., 2020. Convergence of energy productivity in Australian states and territories: Determinants and forecasts. Energy Economics 85, 104538.

Bhattacharya, M., Inekwe, J.N., Sadorsky, P., Saha, A., 2018. Convergence of energy productivity across Indian states and territories. Energy Economics 74, 427-440.

Bilgili, F., Ulucak, R., 2018. Is there deterministic, stochastic and/or club convergence in ecological footprint indicator among G20 countries? Environmental Science and Pollution Research 25 (35), 35404-35419.

Brock, W.A., Taylor, M.S. 2004. Economic growth and the environment: a review of theory and empirics. National Bureau of Economic Research. Working Paper 10854.

Burke, P. J., 2010. Income, resources, and electricity mix. Energy Economics 32 (3), 616–626.

Burnett, J.W., 2016. Club convergence and clustering of U.S. energy related CO2 emissions. Resource and Energy Economics 46, 62-84.

Butnaru, G.I., Haller, A.P., Clipa, R.I., Stefanica, M., Ifrim, M. 2020. The nexus between convergence of conventional and renewable energy consumption in the present European Union states. Explorative study on parametric and semi-parametric methods. Energies 13, 5272.

Camarero, M., Picazo-Tadeo, A.J., Tamarit, C., 2013. Are the determinants of $CO_2$ emissions converging among OECD countries? Economics Letters 118, 159-162.

Carley, S., 2009. State renewable energy electricity policies: An empirical evaluation of effectiveness. Energy Policy 37(8), 3071-3081.

Churchill, S.A., Inekwe, J., Ivanovski, K., 2020. Stochastic convergence in per capita $CO_2$ emissions: Evidence from emerging economies, 1921–2014. Energy Economics 86, 104659.

da Silva, P.P., Cerqueira, P.A., Ogbe, W., 2018. Determinants of renewable energy growth in Sub-Saharan Africa: Evidence from panel ARDL. Energy 156, 45-54.

Demir, C., Cergibozan, R. 2020. Does alternative energy usage converge across OECD countries? Renewable Energy 146, 559-567.

Emir, F., Balcilar, M., Shahbaz, M., 2019. Inequality in carbon intensity in EU-28: analysis based on club convergence. Environmental Science and Pollution Research 26(4), 3308-3319.

Erdogan, S., Okumus, I., 2021. Stochastic and club convergence of ecological footprint: an empirical analysis for different income group of countries. Ecological Indicators 121, 107123.

EU. Directive (EU) 2009/28/EC of the European Parliament and of the Council of 23 April 2009 on the promotion of the use of energy from renewable sources and amending and subsequently repealing Directives 2001/77/EC and 2003/30/EC (Renewable Energy Directive I, RED I). Official Journal of the European Union 140/16.

EU. Directive (EU) 2015/1513 of the European Parliament and of the Council of 9 September 2015 amending Directive 98/70/EC relating to the quality of petrol and diesel fuels and amending Directive 2009/28/EC on the promotion of the use of energy from renewable sources. Official Journal of the European Union 239/1.

EU. Directive (EU) 2018/2001/EC of the European Parliament and of the Council of 11 December 2018 on the promotion of the use of energy from renewable sources (Renewable Energy Directive II, RED II). Official Journal of the European Union 328/82.

European Commission, 2019. Communication from the Commission to the European Parliament, the European Council, the Council, the European Economic and Social Committee and the Committee of the Regions. The European Green Deal. December 11, 2019.

European Commission, 2020. Proposal for a regulation of the European Parliament and of the Council establishing a Recovery and Resilience Facility. May 28, 2020.

European Commission, 2020. Report from the Commission to to the European Parliament, the Council, the European Economic and Social Committee and the Committee of the Regions. Renewable Energy Progress Report. October 14, 2020.

European Commission. Eurostat. "SHARES Tool Manual". Directorate E: Sectoral and regional statistics Unit E.5: Energy). https://ec.europa.eu/eurostat/documents/38154/4956088/SHARES+tool+manual-2019.pdf/8822f775-121e-f73e-3040-93c01f6ebd3e

Filip, O., Janda, K., Kristoufek, L. and Ziberman, D., 2019. Food versus fuel: An updated and expanded evidence. Energy Economics 82, 152-166.

Gan, J.B., Smith, C.T., 2011. Drivers for renewable energy: A comparison among OECD countries. Biomass & Bioenergy 35(11), 4497-4503.

Grossman, G.M., Krueger, A.B., 1991. Environmental impacts of a North American free trade agreement. National Bureau of Economic Research. Working Paper Series nº 3914.

Haider, S., Akram, V., 2019a. Club convergence of per capita carbon emission: global insight from disaggregated level data. Environmental Science Pollution Research 26, 11074–11086.

Haider, S., Akram, V., 2019b. Club convergence analysis of ecological and carbon footprint: evidence from a cros-country analysis. Carbon Management 10, 451-463.

Herrerias, M.J., 2013. The environmental convergence hypothesis: carbon dioxide emissions according to the source of energy. Energy Policy 61, 1140-1150.

Ivanovski, K., Churchill, S.A., 2020. Convergence and determinants of greenhouse gas emissions in Australia: a regional analysis. Energy Economics 92, 104971.

Ivanovski, K., Churchill, S.A., Smyth, R., 2018. A club convergence analysis of per capita energy consumption across Australian regions and sectors. Energy Economics 76, 519–531.

Kasman, A., Kasman, S., 2020. Convergence of renewable energy consumption in the EU-15: evidence from stochastic and club convergence tests. Environmental Science and Pollution Research 27, 5901–5911.

Kim, J., Park, K., 2016. Financial development and deployment of renewable energy technologies. Energy Economics 59, 238-250.

Kim, Y.S. 2015. Electricity consumption and economic development: are countries converging to a common trend?, Energy Economics 49, 192-202.

List, J.A., 1999. Have air pollutant emissions converged among US regions? Evidence from unit root tests. Southern Economic Journal 66, 144-155.

Marques, A. C., Fuinhas, J.A., 2012. Are public policies towards renewables successful? Evidence from European countries. Renewable Energy 44, 109-118.

Morales-Lage, R., Bengoechea-Morancho, A., Camarero, M., Martinez-Zarzoso, I., 2019. Club convergence of sectoral CO2 emissions in the European Union. Energy Policy 135, 111019.

Ohler, A.M., 2015. Factors affecting the rise of renewable energy in the U.S.: Concern over environmental quality or rising unemployment? The Energy Journal 36 (2), 97-115.

Panopoulou, E., Pantelidis, T., 2009. Club Convergence in Carbon Dioxide Emissions. Environmental and Resource Economics 44, 47-70.

Payne J.E., Vizek M., Lee J., 2017. Is there convergence in per capita renewable energy consumption across US states? Evidence from LM and RALS-LM unit root tests with breaks. Renewable and Sustainable Energy Reviews 70, 715-728.

Pfeiffer, B., Mulder, P., 2013. Explaining the diffusion of renewable energy technology in developing countries. Energy Economics 40, 285-296.

Phillips, P.C.B., Sul, D., 2007. Transition modeling and econometric convergence tests. Econometrica 75, 1771–1855.

Phillips, P.C.B., Sul, D., 2009. Economic transition and growth. Journal of Applied Econometrics 24, 1153-1185.

Presno, M.J., Landajo, M., Fernández González, P., 2018. Stochastic convergence in per capita $CO_2$ emissions. An approach from nonlinear stationarity analysis. Energy Economics 70, 563-581.

Presno, M.J., Landajo, M., Fernández González, P., 2021. GHG emissions in the EU-28. A multilevel club convergence study of the Emission Trading System and the Effort Sharing Decision mechanisms. Sustainable Production and Consumption 27, 998-1009.

Reboredo, J.C., 2015. Renewable energy contribution to the energy supply: is there convergence across countries? Renewable and Sustainable Energy Reviews 45, 290-295.

Sadorsky, P., 2009a. Renewable energy consumption and income in emerging economies. Energy Policy 37(10), 4021-4028.

Sadorsky, P., 2009b. Renewable energy consumption, $CO_2$ emissions and oil prices in the G7 countries. Energy Economics 31(3), 456-462.

Solarin, S.A., Gil-Alana, L.A., Al-Mulali, U., 2018. Stochastic convergence of renewable energy consumption in OECD countries: a fractional integration approach. Environmental Science Pollution Research 25, 17289-17299.

Solarin, S.A., Tiwari, A.K., Bello, M.O., 2019. A multi-country convergence analysis of ecological footprint and its components. Sustainable Cities and Society 46, 101422.

Stern, D. 2017. The environmental Kuznets curve after 25 years. Journal of Bioeconomics. 19(1), 7-28.

Strazicich, M.C., List, J.A., 2003. Are $CO_2$ emission levels converging among industrial countries? Environmental and Resource Economics 24, 263-271.

Ulucak, R., Apergis, N., 2018. Does convergence really matter for the environment? An application based on club convergence and on ecological footprint concept for the EU. Environmental Science and Policy 80, 21-27.

Wang, Y., Zhang, P., Huang, D., Cai, C., 2014. Convergence behavior of carbon dioxide emissions in China. Economic Modelling 43, 75-80.

Yu, S., Hu, X., Fan, J., 2018. Convergence of carbon emissions intensity across Chinese industrial sectors. Journal of Cleaner Production 194, 179-192.

# TABLES AND FIGURES

**Table 1. Convergence in RES indicators (2004-2018).**

|  |  | Initial classification | | | Test of club merger | | | Transition | | |
|---|---|---|---|---|---|---|---|---|---|---|
|  |  | $t_{\hat{b}}$ | $\hat{b}$ | $\hat{\alpha}$ |  | $t_{\hat{b}}$ | $\hat{b}$ |  | $t_{\hat{b}}$ | $\hat{b}$ |
| **PANEL A** | | | | | | | | | | |
| RES |  | -14.817* | -0.229 | -0.114 | | | | | | |
|  | **Club 1**. SE, FI, LV, DK, AT, PT | 0.604 | 0.031 | 0.015 | - | - | - | DK, AT, PT+EE, HR, LT, RO, SI, BG, EL, IT, ES, FR, DE | -3.003* | -0.197 |
|  | **Club 2**. EE, HR, LT, RO, SI, BG, EL, IT, ES, FR, DE, CZ, CY, HU, SK, PL, IE, UK, BE, LU, MT, NL | 1.635 | 0.143 | 0.071 | | | | | | |
| RES-T | | -4.366* | -0.770 | -0.385 | | | | | | |
|  | **Club 1**. SE, FI, BG, MT, DK, RO | 2.410 | 0.528 | 0.264 | - | - | - | MT, DK, RO+AT, NL, FR, PT, DE, HU, IT, IE, SK, ES, BE | 4.305 | 2.668 |
|  | **Club 2**. AT, NL, FR, PT, DE, HU, IT, IE, SK, ES, BE, CZ, LU, UK, PL, SI, LV, LT, HR, EL, EE, CY | 0.789 | 0.353 | 0.176 | | | | | | |
| RES-HC | | -12.062* | -0.447 | -0.223 | | | | | | |
|  | **Club 1**. SE, LV, FI, EE, DK, LT, CY, MT | 2.081 | 0.344 | 0.172 | Club1+Club2 | -2.015* | | DK, LT, CY, MT+PT, HR, AT, BG, | 1.045 | 0.306 |
|  | **Club 2**. PT, HR, AT, BG, SI, EL, RO, FR, CZ | -0.683 | -0.042 | -0.021 | Club2+Club3 | -21.421* | -0.166 | SI, EL, RO, FR, CZ+IT, HU, ES, PL, DE, SK | -7.315* | -0.447 |
|  | **Club 3**. IT, HU, ES, PL, DE, SK, LU, BE, UK, IE, NL | -0.335 | -0.063 | -0.031 | | -0.412 | | | | |
| RES-E | | 3.329 | 0.055 | 0.027 | | | | | | |
| **PANEL B** | | | | | | | | | | |
| RES (objective) | | 3.460 | 0.678 | 0.339 | | | | | | |

*denotes rejection of the null of convergence at 5% significance level.

Belgium (BE), Bulgaria (BG), Czech Republic (CZ), Denmark (DK), Germany (DE), Estonia (EE), Ireland (IE), Greece (EL), Spain (ES), France (FR), Croatia (HR), Italy (IT), Cyprus (CY), Latvia (LV), Lithuania (LT), Luxembourg (LU), Hungary (HU), Malta (MT), Netherlands (NL), Austria (AT), Poland (PL), Portugal (PT), Romania (RO), Slovenia (SI), Slovakia (SK), Finland (FI), Sweden (SE), United Kingdom (UK).

**Table 2: Explicative variables in the probit model for club membership.**

| GDPCAP | Real GDP per capita (in euro, average for the 2010-2018 period). |
|---|---|
| SQ_GDPCAP | Square of GDPCAP. |
| ENVEXPGDP | National expenditure (total economy) on environmental protection divided by nominal GDP (%, average ratio for the 2014-2016 period). |
| ENIMPDEP | Energy import dependency (%, average for the 2009-2018 period). |
| NUCLENCAP | Nuclear enrichment capacity (in tSWU, average for the 2010-2018 period). |

Source: Eurostat.

**Table 3: Results of the probit model for club membership.**

|  | *Coefficient* | *Robust s.e.* | *z-stat* | *p-value* |
|---|---|---|---|---|
| Constant | 1.49391 | 1.05893 | 1.411 | 0.1583 |
| GDPCAP** | -0.000110787 | 5.26560e-05 | -2.104 | 0.0354 |
| SQ_GDPCAP** | 9.68343e-010 | 4.92566e-010 | 1.966 | 0.0493 |
| ENVEXPGDP** | -0.0963451 | 0.0458660 | -2.101 | 0.0357 |
| ENIMPDEP** | 0.0303434 | 0.0120250 | 2.523 | 0.0116 |
| NUCLENCAP*** | 0.00430606 | 0.00156178 | 2.757 | 0.0058 |
| Mean of dep. var. | 0.785714 | S.e. of dep. var. | | 0.417855 |
| McFadden's R-squared | 0.476376 | Schwarz's criterion | | 35.22884 |
| Log-likelihood | -7.617805 | Akaike's criterion | | 27.23561 |
| No. of cases "correctly predicted" | 26 (92.9%) | | | |
| LR test | | | | |
| Chisq. Stat. | 13.8609 | p-value | | 0.0021 |

** and *** denote, respectively, significance at 5% and 1% levels.

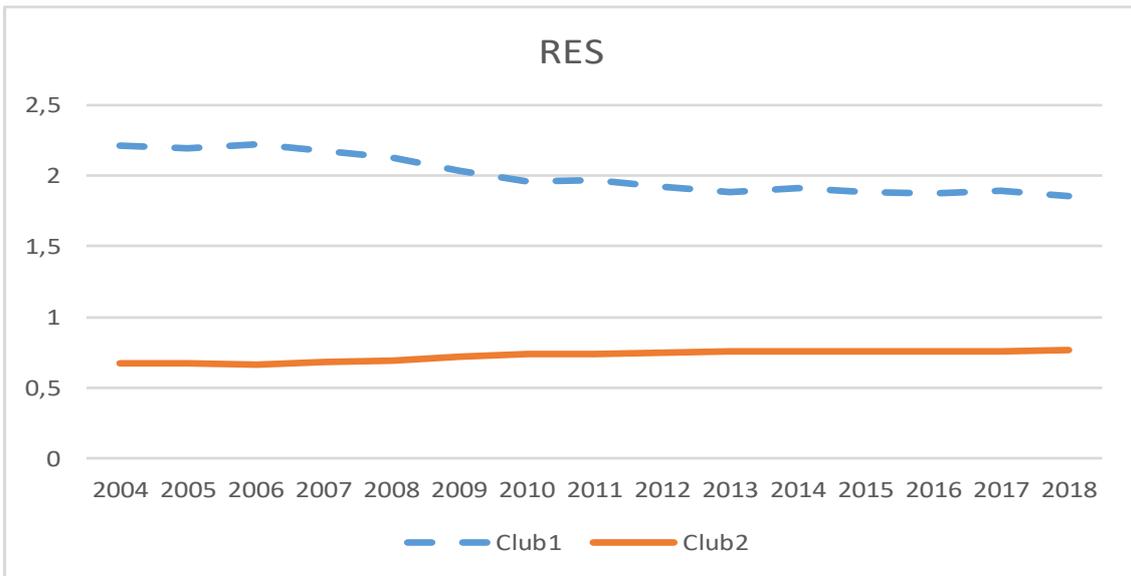
**Figure 1. RES transition paths.**

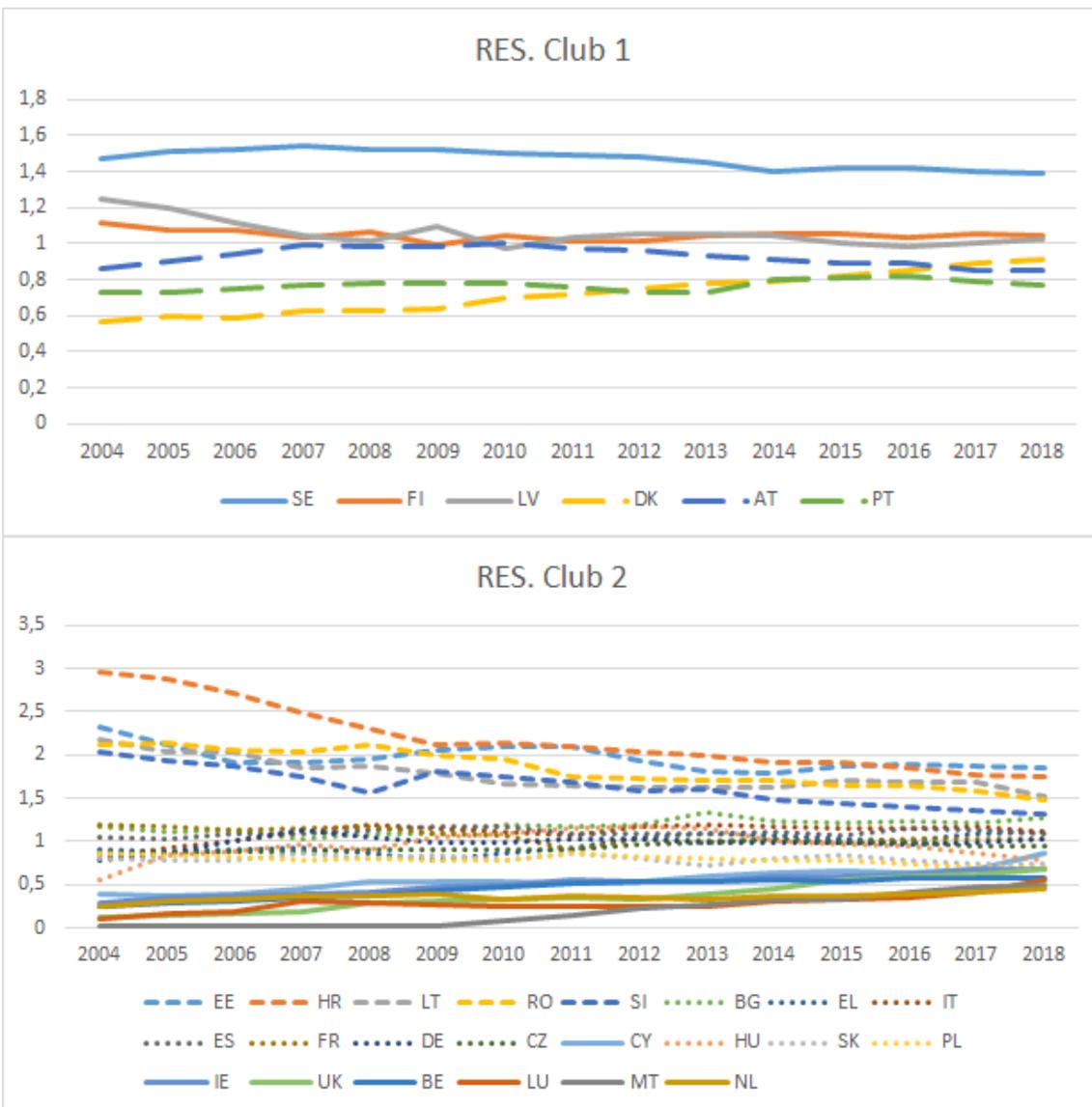
**Figure 2. RES transition paths (with respect to the mean of the club).**

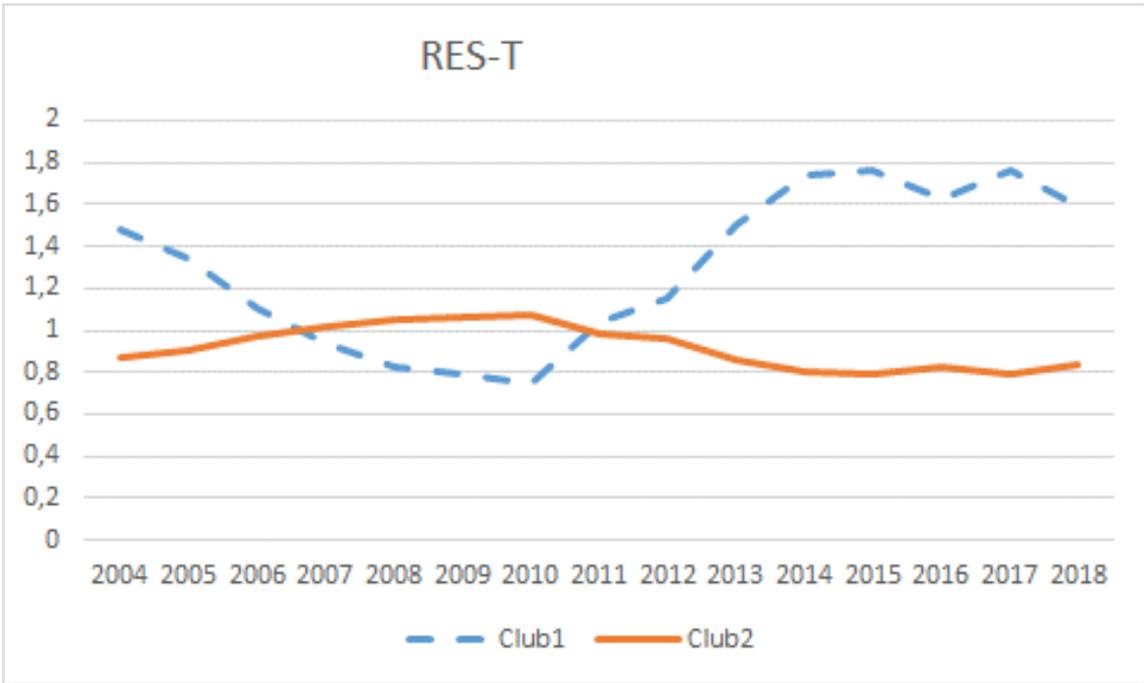

**Figure 3.** Transition paths in RES-T.

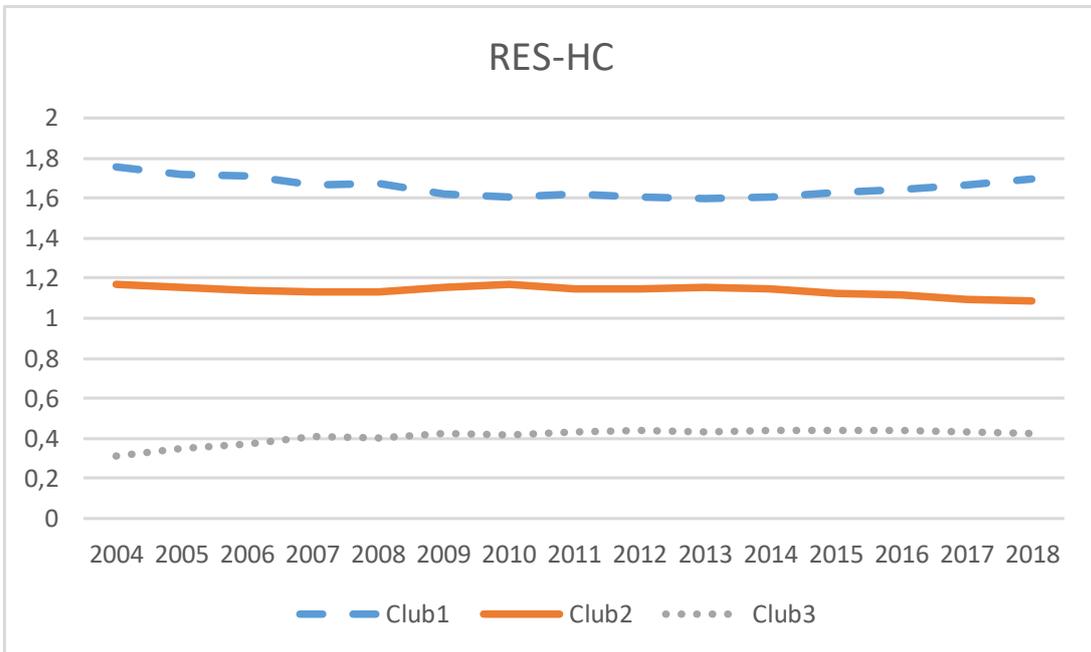

**Figure 4.** Transition paths in RES-HC.

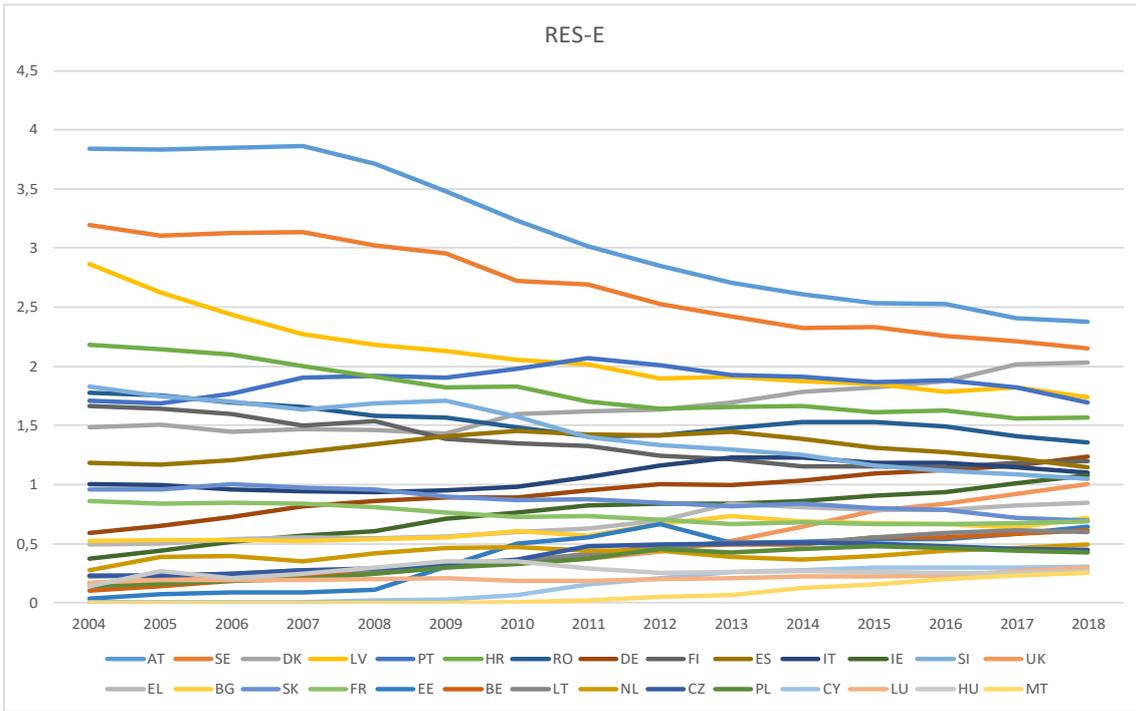

**Figure 5.** Transition paths in RES-E.